\documentclass[Journal]{IEEEtran}
\usepackage{amsmath,amsthm}
\usepackage{cite}
\usepackage{graphicx}
\usepackage{epstopdf}
\usepackage{color}
\usepackage{amsfonts,amsmath,amssymb}
\usepackage{cite}
\usepackage{graphicx}
\usepackage{url}
\usepackage{bm}
\usepackage{bbm}
\usepackage{amssymb}
\usepackage{bbold} 
\def\BibTeX{{\rm B\kern-.05em{\sc i\kern-.025em b}\kern-.08em T\kern-.1667em\lower.7ex\hbox{E}\kern-.125emX}}
\usepackage{fancyhdr}
\usepackage{lipsum}

\usepackage{caption}
\usepackage{subcaption}
\DeclareCaptionFont{mysize}{\fontsize{8}{9.6}\selectfont}

\newcommand{\ee}{\mathrm{e}}
\newcommand{\ii}{\mathrm{i}}
\newcommand{\bal}{\begin{align}}
\newcommand{\eal}{\end{align}}
\newcommand{\beq}{\begin{equation}}
\newcommand{\eeq}{\end{equation}}
\newcommand{\ket}[1]{| #1 \rangle}

\begin{document}
	\bstctlcite{IEEEexample:BSTcontrol}

\title{Teleportation of a Schr\"odinger's-Cat State via Satellite-based Quantum Communications}
\author{
    \IEEEauthorblockN{Hung Do, Robert Malaney and Jonathan Green \thanks{Hung Do and Robert Malaney are at the University of New South Wales, Sydney, NSW, Australia. Jonathan Green is at Northrop Grumman Corporation, San Diego, California, USA. }}
}

\maketitle
\thispagestyle{fancy}
\renewcommand{\headrulewidth}{0pt}

\begin{abstract}
 
The Schr\"odinger's-cat state is created from the macroscopic superposition of coherent states and is well-known to be a useful resource for quantum information processing protocols. 
In order to extend such protocols to a global scale, we study the continuous variable (CV) teleportation of the cat state via a satellite in low-Earth-orbit. 
Past studies have shown that the quantum character of the cat state can be preserved after CV teleportation, even when taking into account the detector efficiency. However, the channel transmission loss has not been taken into consideration. Traditionally, optical fibers with fixed attenuation are used as teleportation channels. Our results show that in such a setup, the quantum character of the cat state is lost after 5dB of channel loss. We then investigate the free-space channel between the Earth and a satellite, where the loss is caused by atmospheric turbulence. For a down-link channel of less than 500km and 30dB of loss, we find that the teleported state preserves higher fidelity relative to a fixed attenuation channel. The results in this work will be important for  deployments over Earth-Satellite channels of protocols dependent on cat-state qubits.

\end{abstract}

\section{Introduction}


The Schr\"odinger's-cat state is of fundamental interest because it represents quantum superposition between two macroscopic states. Experimentally, the cat state can be created from the macroscopic superposition of coherent states. Such a superposition is known to be useful for universal quantum computing using continuous variables (CV), where the quantum information is carried by the quadratures of the optical field \cite{lund2019fault}. CV protocols use homodyne (or heterodyne) detectors which are faster and more efficient (as compared to discrete variable (DV) protocols, where the quantum information is carried in the polarization or the photon-number of single photons).

\begin{figure}[h!]
    \centering
        \includegraphics[width=\linewidth]{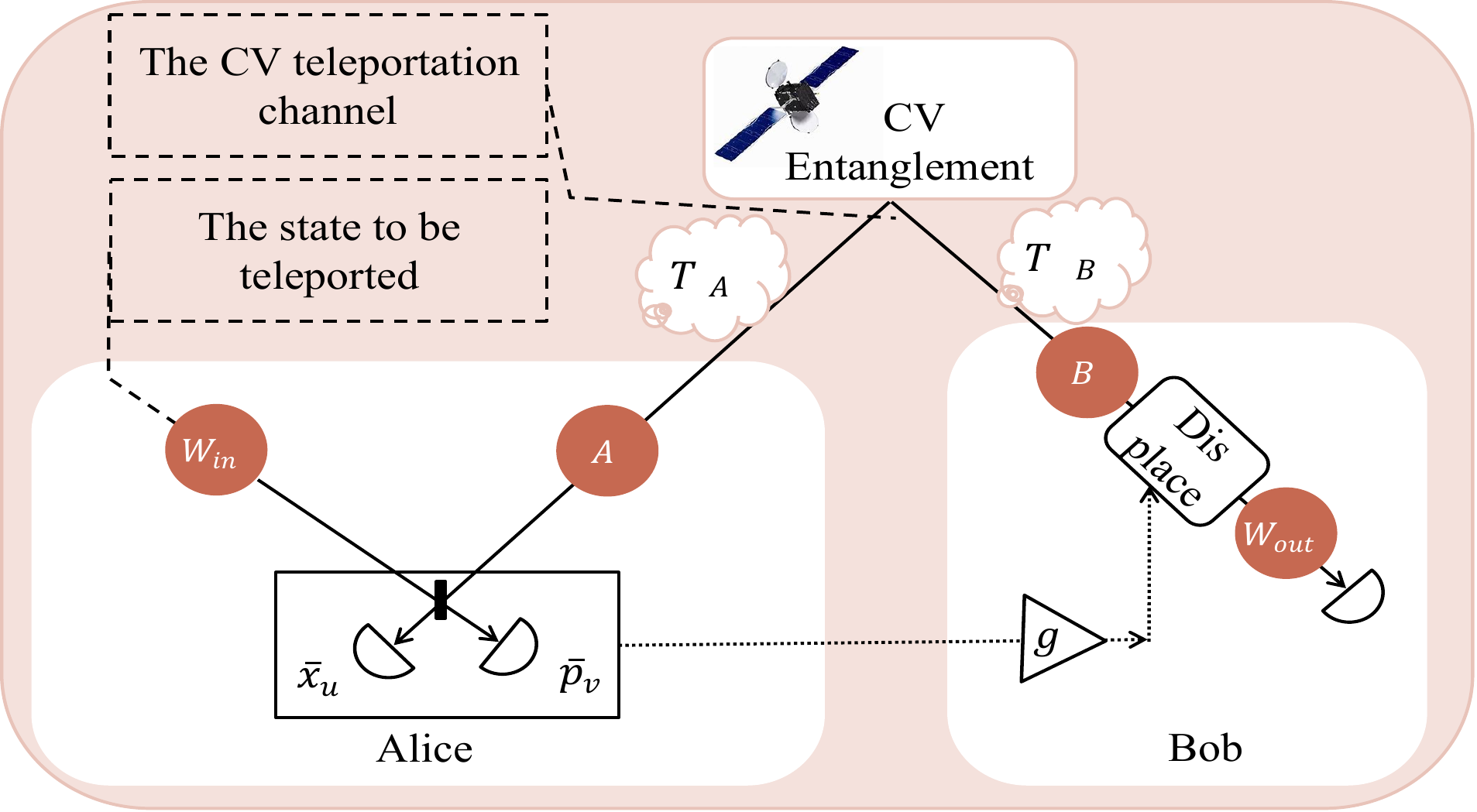}
		\caption{ In the hybrid scheme, the CV entanglement $A-B$ is used as a teleportation channel. The channels have transmissivity $T_A$ and $T_B$. The input state $W_{in}$ is teleported from Alice to Bob, resulting in the output $W_{out}$.}
		\label{setup}
\end{figure}

Among the different quantum communication protocols, teleportation is one of the most useful. In many cases, information is transferred through optical fibers, which exhibits fixed attenuation. 
Since the loss in optical fiber scales exponentially with distance, terrestrial quantum communication has been limited to a few hundred kilometer. Quantum signal cannot be further amplified due to the no-cloning theorem \cite{park1970concept,dieks1982communication,wootters1982noclone}. In contrast, for a satellite in low orbit (400 to 600km from Earth), the channel is only affected by atmospheric turbulence. In fact, for a down-link channel of less than 500km above the ground, the average attenuation was demonstrated to be around 30dB \cite{yin2017satellite}. In this work, we use CV entanglement distributed from a satellite as a CV teleportation channel. We then use this channel to distribute a Schr\"odinger-cat state. Such a protocol has the potential to extend quantum information protocols over a global scale.

The satellite-Earth channel is a free-space fading channel, where the loss is caused by diffraction, absorption/scattering, and atmospheric turbulence \cite{hosseinidehaj2018satellite}. The most significant loss comes from the turbulence, which could be characterized by the Rytov parameter $\sigma_R$. This parameter is a function of the atmospheric refraction index structure constant parameter $C_n^2$, which can be calculated from the wind velocity and the altitude. Large turbulent eddies can cause beam-wandering, while smaller eddies can induce beam broadening. 
The channel attenuation has a probabilistic nature and can be simulated by different mathematical models.
In the beam-wandering model, the beam profile is modeled by a circular Gaussian shape. The turbulence causes the beam center to fluctuate around the center of the receiver aperture, resulting in a misalignment which reduces the transmissivity  of the beam \cite{vasylyev2012toward}. 
Recently, a new fading model was introduced taking into account additional deformation and broadening effects. In this paper, we call this model the elliptic model. In the elliptic model, the beam profile at the receiver aperture is broadened and has an elliptic shape. There is fluctuations not only in the beam center but also in the widths of the elliptic profile \cite{vasylyev2016atmospheric}. 

Past studies have shown that the quantum character of the cat state can be preserved after CV teleportation \cite{lee2011teleportation}, however, the channel transmission loss has not been taken into consideration.  
Our main contributions in this paper are:
\begin{itemize}
\item We derive a mathematical model detailing how channel attenuation affects the quality of the CV teleportation channel, as well as the quality of the teleportation outcome.

\item  For fixed attenuation we determine the loss conditions under which the cat state can be preserved. 

\item We adopt the free-space fading channel between
the Earth and a satellite for a typical down-link
channel and determine whether the teleported cat
state retains higher fidelity in such channels relative to a fixed attenuation channel.
\end{itemize}

The structure of the remainder of this paper is as follows. Section \ref{teleportation} studies the mathematical model for CV teleportation. Section \ref{channelmodel} studies different models of atmospheric turbulence.
Section \ref{simulation} shows our simulation, while section \ref{conclusion} summarizes our findings and discusses future work.

\section{Teleportation by an attenuated CV teleportation channel}
\label{teleportation}
\subsection{Attenuation on a CV teleportation channel}
Among different CV entanglement resources, the two-mode squeezed vacuum (TMSV) state is widely used, in part, because it has a Gaussian form. The Wigner function of a TMSV state can be written as \cite{braustein2005QIwithCV}
\begin{align}
&W_{TMSV}(\zeta_A, \zeta_B) = \frac{4}{\pi^2}\exp\{\nonumber\\
&-e^{-2r}\left[(x_A+x_B)^2+(p_A-p_B)^2\right]  \nonumber\\
&-e^{+2r}\left[(x_A-x_B)^2+(p_A+p_B)^2\right]\},
\end{align}
where $r$ is the squeezing parameter, $\zeta_i = (x_i, p_i)$, $x_i$ and $p_i$ are the quadrature values, with subscript $i \in \{A,B\}$ denoting the corresponding modes. 
When $r\rightarrow \infty$,  $W_{TMSV}(\zeta_A, \zeta_B)  \propto \delta(x_A-x_B)\delta(p_A+p_B)$, which represents a maximally entangled EPR state.

For a TMSV state, when the modes are sent through two channels with different transmissivities $T_A$ and $T_B$ (see Fig. \ref{setup}), our calculation shows that the state becomes
\begin{align}
&W_{TMSV}^{(T_A, T_B)}(\zeta_A, \zeta_B) = \frac{4}{\pi^2 \tau}\exp\{\nonumber\\
&-\frac{e^{-2r}}{\tau}\left[\left(x_A\sqrt{T_B}+x_B\sqrt{T_A}\right)^2+\left(p_A\sqrt{T_B}-p_B\sqrt{T_A}\right)^2\right]  \nonumber\\
&-\frac{e^{+2r}}{\tau}\left[\left(x_A\sqrt{T_B}-x_B\sqrt{T_A}\right)^2+\left(p_A\sqrt{T_B}+p_B\sqrt{T_A}\right)^2\right]\nonumber\\
&-\frac{2}{\tau}\left[(1-T_B)\left(x_A^2+p_A^2\right)
+(1-T_A)\left(x_B^2 + p_B^2\right)\right]\},
\label{TMSV_T}
\end{align}
where $\tau$ is given by
\beq
\tau = 1+\left[\cosh (2r)-1\right](T_A+T_B -2 T_A T_B).
\eeq
In this work, an attenuated CV entangled state is used as the CV teleportation channel (see Fig. \ref{setup}).

\subsection{Teleportation of an arbitrary mode}
\label{cv_teleportation}
Let $W_{in}(\zeta)$ be the input state (Fig. \ref{setup}). After CV teleportation with gain $g$, the output Wigner function becomes a Gaussian 
convolution of the input, followed by a rescaling of $\zeta_{out} \rightarrow \zeta_{out}/g$ in phase space.
A general Gaussian function with variance $V$ can be written as
\beq
G_V(\zeta) =\frac{1}{\pi V}\iint{dx dp \exp\left(-\frac{x^2 + p^2}{V}\right)},
\label{Gaussian}
\eeq
where $\zeta$ is $(x, p)$. We have
\begin{align}
W_{out}(\zeta_{out}) &= \frac{1}{g^2}\left[W_{in}*G_V\right](\frac{\zeta_{out}}{g})\nonumber\\
&= \frac{1}{g^2}\iint dx dp W_{in}(\zeta) G_V\left(\frac{\zeta_{out}}{g}-\zeta\right),
\label{general_teleportation}
\end{align}
with variance $V$ given by
\begin{align}
V & = \frac{1}{4g^2}\;\,[\;\, e^{+2r}(g\sqrt{T_A}-\sqrt{T_B})^2 \nonumber\\
&\qquad\quad +e^{-2r}(g\sqrt{T_A}+\sqrt{T_B})^2 \nonumber\\
&\qquad\quad +2g^2\;\, (1-T_A)+2(1-T_B)\; ].
\label{sigma_general}
\end{align}
When $T_A = T_B =1$, we have \cite{liu2003improving}
\beq
V = \frac{e^{2r}(1-g)^2+e^{-2r}(1+g)^2}{4g^2}.
\eeq
When we further set $g=1$, the variance is simplified to $V = e^{-2r}$, so $W_{out}(\zeta_{out})= W_{in}*G_{e^{-2r}}(\zeta_{out})$ \cite{vaidman1994teleportation, braustein1998teleportationCV, braustein2005QIwithCV}. 

\newcommand{\alsq}{|\alpha_0|^2}

\subsection{Teleportation of a Schr\"odinger cat state}

In this section, we will study the specific case where the input is a Schr\"odinger's-cat state. The cat state is formed by a superposition of coherent states \cite{Girish2013QuantumOptics}
\beq
\ket{\alpha}_c = N^{-1} \left(\ket{\alpha_0} + \ee^{\ii\phi}\ket{-\alpha_0} \right),
\label{cat_state_eqn}
\eeq
where $\alpha_0$ is a complex number, $\phi$ is the phase and $N$ is the normalization constant
\beq
N = \sqrt{2+ 2\ee^{-2|\alpha_0|^2}\cos\phi}.
\eeq
Let $\alpha = x + \ii p$, then the Wigner function of the cat state is 
\begin{align}
W_c(\alpha) = \frac{2N^{-2}}{\pi}\{
&\exp\left( -2|\alpha - \alpha_0|^2 \right)\nonumber\\
&+ \exp\left( -2|\alpha + \alpha_0|^2 \right) \nonumber\\
&+ 2\ee^{-2|\alpha|^2}\cos \left[ \phi + 4 \textrm{Im}(\alpha_0*\alpha) \right] 
\}.
\label{cat_wigner}
\end{align}
When $W_c(\alpha)$ is used as the input state $W_{in}(\zeta)$, with $\zeta = (x,p)$, the output of the CV teleportation is calculated by Eq. (\ref{general_teleportation}). When the CV teleportation channels has transmissivities $T_A = T_B = T$ and the optimal gain is unity, the Gaussian variance $V$ in Eq. (\ref{sigma_general}) becomes
\beq
V = T\ee^{-2r} + (1-T).
\label{sigma_T}
\eeq
When taking into account the detector amplitude-efficiency $\eta$, the Gaussian variance is modified to $V' = V + \frac{1 - \eta^2}{\eta^2}$. 

In order to quantify the quality of the output state as compared to input state, we use the entanglement fidelity \cite{schumacher1996sending}. For the Schr\"odinger's-cat state defined in Eq. (\ref{cat_state_eqn}), the fidelity was given in \cite{braustein1998teleportationCV}
\beq
F = \frac{1}{1 + V'} - 
\frac{1+ \ee^{-4\alsq} - \exp{\frac{-4V'\alsq}{1 + V'}} 
- \exp{\frac{-4\alsq}{1 + V'}} }
{2(1+V')(1 + \ee^{-2\alsq}\cos\phi )^2}.
\label{fidelity}
\eeq


\section{Earth-satellite Channel}
\label{channelmodel}

\subsection{Atmospheric turbulence}

The level of atmospheric turbulence is characterized by the Rytov parameter \cite{vasylyev2016atmospheric}
\beq
	\sigma_R^2 = 1.23C_n^2k^{7/6}L^{11/6},
\eeq
where $k$ is the wave number of the light mode and $L$ is the altitude of the satellite. The refraction index structure constant parameter $C_n^2$ can be calculated from the altitude $h$ and wind velocity $v$ \cite{guo2018submarine}

\begin{eqnarray}
	C_n^2(h) &=& 0.00594(v/27)^2(h\times 10^{-5})^{10} e^{-h/1000}\nonumber\\
	&\,& + 2.7\times 10^{-16}e^{-h/1500}+ Ae^{-h/100},
\end{eqnarray}
where $A = C_n^2(h=0)$ is the refraction index constant structure parameter at ground level, in units of m$^{-2/3}$. The mean value of $A$ according to WRF model is 9.6x10$^{-14}$m$^{-2/3}$. 

\subsection{Beam-wandering model}
Beam wandering causes the beam center to be displaced from the aperture center. On the aperture plane, the position of the beam centroid relative to the aperture center can be modeled by two random variables $(x,y)$ following two independent Gaussian probability distributions with variance $\sigma^2$
\beq
p(x, y) = p(x)p(y)=\frac{1}{2\pi \sigma^2} \exp \left( - \frac{x^2 + y^2}{2 \sigma^2}\right).
\eeq
Thus, the displacement $d = \sqrt(x^2 + y^2)$ follows the Rice distribution:
\beq
	p(d) = \frac{d}{\sigma^2}\exp \left( - \frac{d^2}{2 \sigma^2} \right).
	\label{eq:pr0}
\eeq
\begin{figure}[h!]
    \centering
		\begin{subfigure}[b]{0.46\linewidth}
        \includegraphics[width=\linewidth]{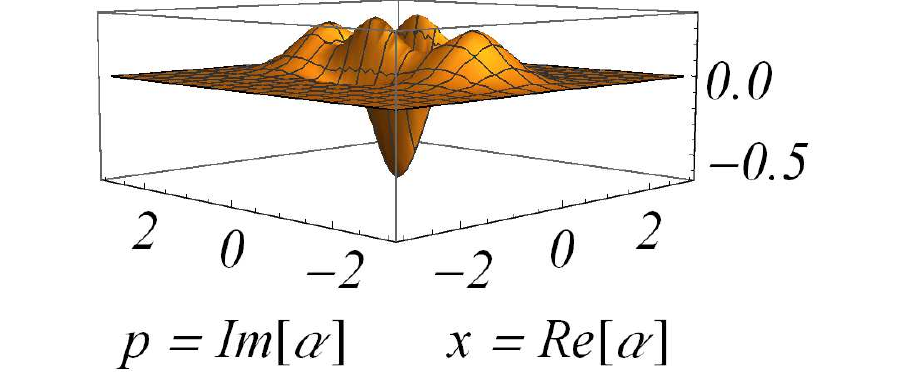}
        \caption{Input cat state.}
    \end{subfigure}
		\\ \vspace{7mm}
~
    \begin{subfigure}[b]{0.48\linewidth}
        \includegraphics[width=\linewidth]{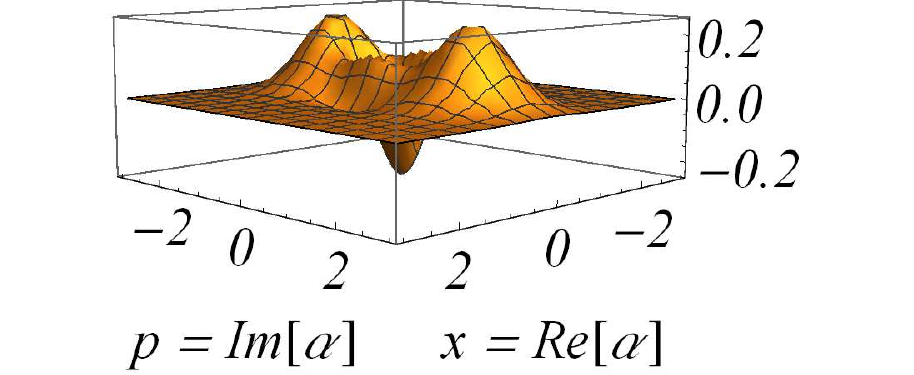}
        \caption{Teleported cat state, no loss, $\eta^2 =1$.}
    \end{subfigure}
		\qquad
    \begin{subfigure}[b]{0.39\linewidth}
        \includegraphics[width=\linewidth]{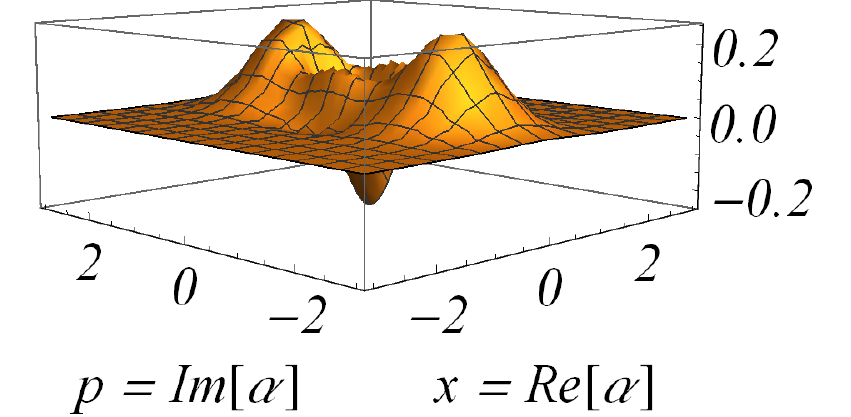}
        \caption{Teleported cat state, no loss, $\eta^2 =0.99$.}
    \end{subfigure}
		\\\vspace{7mm}
    ~ 
		\begin{subfigure}[b]{0.44\linewidth}
        \includegraphics[width=\linewidth]{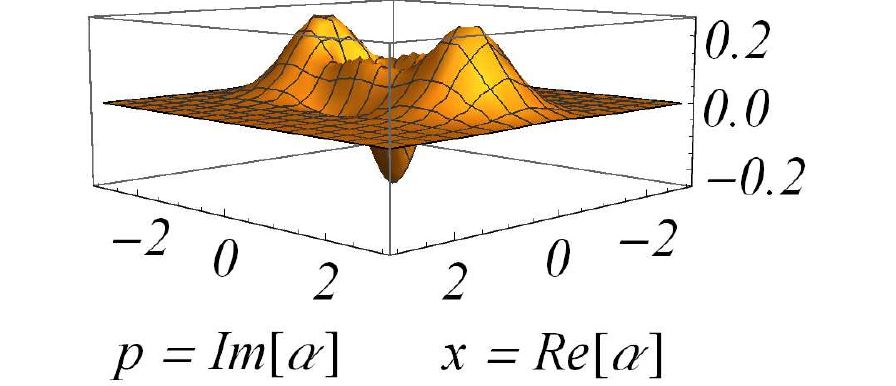}
        \caption{Teleported cat state, 5dB loss, $\eta^2 = 1$.}
    \end{subfigure}
		\qquad
    \begin{subfigure}[b]{0.44\linewidth}
        \includegraphics[width=\linewidth]{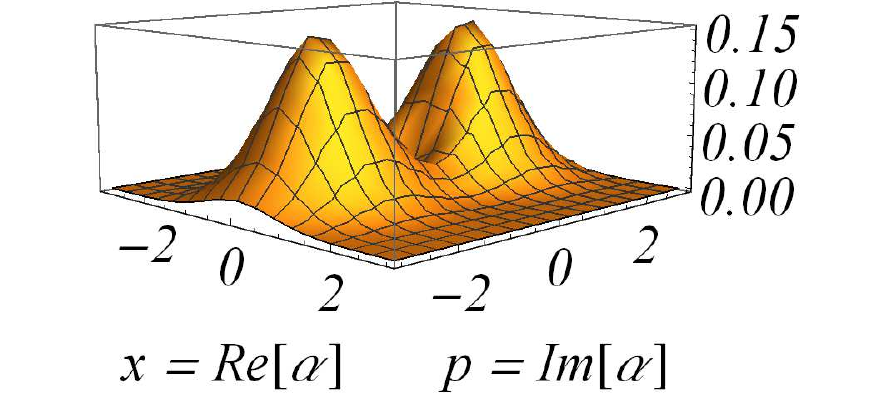}
        \caption{Teleported cat state, 5dB loss, $\eta^2 = 0.99$.}
    \end{subfigure}
    \caption{(a) The Wigner function of a Schr\"odinger cat state, with $\alpha_0 = 1.5 \ii$ and $\phi = \pi$. (b), (c), (d), (e) The Wigner function of the cat state after CV teleportation (with $r$=1.15) is plotted for different values of channel loss and detector amplitude-efficiency ($\eta$).}
\label{cat_states}
\end{figure}

The Rytov parameter $\sigma_R^2$ is related to the beam-wandering variance $\sigma^2$ by \cite{vasylyev2016atmospheric}
\begin{equation}
	\sigma^2 = 0.33 W_0^2 \sigma_R^2 \Omega^{-7/6},
\end{equation}
where
\begin{equation}
	\Omega = \frac{\pi W_0^2}{L \lambda},
\end{equation}
and $W_0$ is the beam width at the transmitter plane. 
Let $W$ be the beam width at the receiver plane, let $a$ be the aperture radius at the receiver, we define the ratios $d_a = d_0/a$ and $W_a = W/a$. The transmittance of the beam is found by integration to find the fraction of energy passing through the aperture. Let $t$ be the transmission coefficient, the channel transmissivity $T=t^2$ can be approximated by
\beq
	T = T_0 \exp\left[ -\left( \frac{d_a}{R_a}\right)^\gamma \right],
	\label{eq:t2_beamwandering}
\eeq
where the parameters are defined as
\begin{eqnarray}
	T_0 &=& 1-\exp\left(-\frac{2}{W_a^2}\right), \label{t0_circular}\\
	\lambda &=& \frac{8}{W_a^2}\times
	\frac{\exp\left(-\frac{4}{W_a^2}\right) I_1\left(\frac{4}{W_a^2}\right)}
		{1-\left(-\frac{4}{W_a^2}\right) I_0\left(\frac{4}{W_a^2}\right)}\nonumber\\
		&\;& \times
		\frac{1}{\ln\left[\frac{2t_0^2}{1-\left(-\frac{4}{W_a^2}\right) I_0\left(\frac{4}{W_a^2}\right)}\right]},\\
	R_a &=& \left[ \ln\left(\frac{2t_0^2}
	{1-\exp\left(-\frac{4}{W_a^2}\right) I_0\left(\frac{4}{W_a^2}\right)}\right)\right]^{-\frac{1}{\lambda}},
\end{eqnarray}
and where $I_0(x)$ and $ I_1(x)$ are the modified Bessel functions. By plugging the probability distribution (Eq. (\ref{eq:pr0})) into the transmissivity (Eq. \ref{eq:t2_beamwandering})), we can derive the probability distribution of the transmission coefficient $t$
\beq
p(t) = \frac{2R_a^2}{\sigma_a^2 \lambda t}
\left(2\ln\frac{t_0}{t}\right)^{\left(\frac{2}{\lambda}-1\right)}
\exp\left[-\frac{R_a^2}{2\sigma_a^2}\left(2\ln\frac{t_0}{t}\right)^\frac{2}{\lambda}\right],
\eeq
where $\sigma_a = \sigma / a$. 
Essentially, the distribution of the transmission coefficient $t$ does not depend on the absolute value of the aperture radius $a$, but only depends on the ratios of $W$ and $\sigma$ to $a$.


\begin{figure*}
	\includegraphics[width=0.85\textwidth]{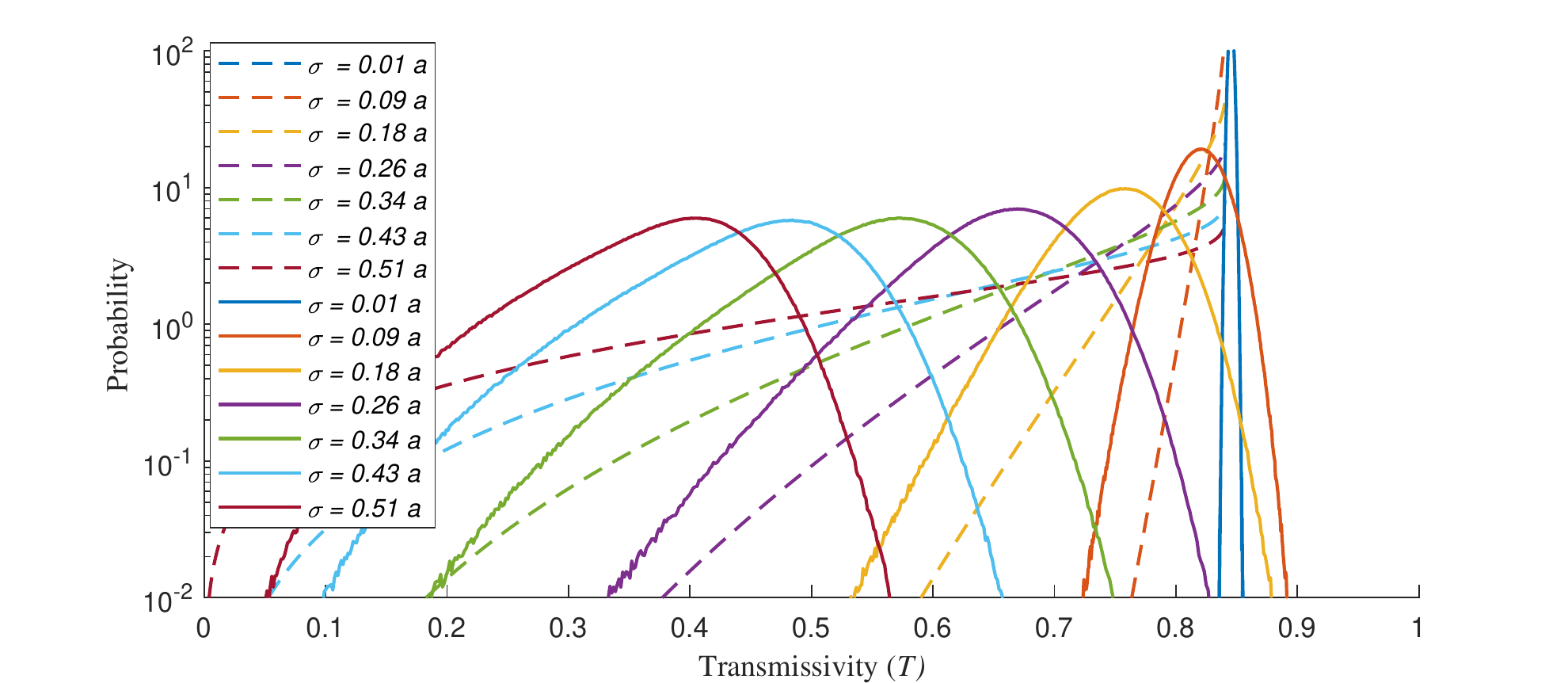}
	\centering
	\caption{For a free-space fading channel, the distribution of channel transmissivity $T$ is plotted with different ratios of the beam-wandering standard deviation ($\sigma$) to the aperture radius ($a$). The dashed lines represent the distributions from the beam-wandering model, while the solid lines represent the distributions from the elliptic model.} 
	\label{circular_vs_elliptical_t_all_sigma_1e7}
\end{figure*}
\subsection{Elliptic model}

The elliptic model takes into account not only the beam-wandering effect, but also beam-broadening, and the non-circular beam distortions caused by turbulence in the Earth's atmosphere \cite{andrews2005laser}\cite{vasylyev2016atmospheric}. Let $x$ and $y$ be the deviation (on the $x$- and $y$-axis, respectively) of the beam centroid from the aperture center, $a$ be the aperture radius of the detector, and $\phi$ be the beam rotation angle, $\phi_0=\tan^{-1}\frac{y}{x}$. 
The effective spot radius of an equivalent Gaussian beam is modelled as
\begin{equation}
\begin{array}{*{20}{l}}	
W_{\rm{eff}}^2(\phi)= & 4a^2\left\lbrace\mathcal{W}\left(\frac{4a^2}{W_1W_2}e^{(a^2/W_1^2)\left[1+2\cos^2(\phi)\right]}\right.\right.\\ &
\times\left.\left. e^{(a^2/W_2^2)\left[1+2\sin^2(\phi)\right]}\right)\right\rbrace^{-1}\ ,
\end{array}
\end{equation}
where $\mathcal{W}(\cdot)$ is the Lambert W function.
The maximal transmissivity at $x=y=0$ (i.e. no deviation) is   
\begin{equation}
\begin{array}{*{30}{l}}
T_{E_0}=
& 1 - I_0\left(a^2\left[\frac{1}{W_1^2}-\frac{1}{W_2^2}\right]\right)e^{-a^2\left(1/W_1^2+1/W_2^2\right)}\\
& - 2\left[1-e^{-(a^2/2\left[1/W_1-1/W_2\right]^2)}\right] \\
&\times\exp\left\lbrace-\left[\frac{\frac{(W_1+W_2)^2}{\left|W_1^2-W_2^2\right|}}{R\left(\frac{1}{W_1}-\frac{1}{W_2}\right)}\right]^{\lambda\left(\frac{1}{W_1}-\frac{1}{W_2}\right)}\right\rbrace
\end{array},
\label{t_0_elliptic}
\end{equation}
where $R(W)$ and $\lambda(W)$ are the scaling and shaping functions
\begin{equation}		
R(W) = \left[\ln\left(2\frac{1-\exp\left[-\frac{1}{2}a^2W^2\right]}{1-\exp\left[-a^2W^2\right]I_0(a^2W^2)}\right)\right]^{-\lambda^{-1}},
\end{equation}	

\begin{equation}
\begin{array}{*{20}{l}}
\lambda(W) = & 2a^2W^2\frac{\exp{[-a^2W^2]}I_1(a^2W^2)}{1-\exp\left[-a^2W^2\right]I_0(a^2W^2)} \\ &
\times \left(2\frac{1-\exp\left[-\frac{1}{2}a^2W^2\right]}{1-\exp\left[-a^2W^2\right]I_0(a^2W^2)}\right),
\end{array}
\label{Cheq5}
\end{equation}
respectively. Here $W_1$ and $W_2$ are the semi-major and semi-minor axis lengths of the ellipse, respectively, and $I_i(\cdot)$ is the modified Bessel function of $i$-th order.
The channel transmissivity $T_E=t_E^2$ is given by
\begin{equation}
T_E = T_{E_0}\exp\left\lbrace-\left[\frac{\sqrt{x^2+y^2}/a}{R(\frac{2}{W_{eff}(\phi-\phi_0)})}\right]^{ \lambda\left(2/W_{eff}(\phi-\phi_0)\right) }\right\rbrace.
\label{Cheq1}
\end{equation}

When considering isotropic turbulence, the rotation angle $\phi$ is evenly distributed. The statistical properties of $\left\lbrace x,y,\theta_1,\theta_2\right\rbrace$ are determined by $W_0$, $\sigma_R^2$, the wavelength of the beam, and the propagation distance. For more details, see supplementary materials in reference \cite{vasylyev2016atmospheric}.


\section{Simulation and results}
\label{simulation}
\subsection{Fixed attenuation channel}
In this subsection, we perform the teleportation of a Schr\"odinger's-cat state following Eqs. (\ref{general_teleportation}), (\ref{cat_wigner}) and (\ref{sigma_T}). The channels are assumed to be symmetric with fixed attenuation $T_A = T_B =T$. The fixed parameters are the squeezing parameter $r$ = 1.15, $\phi = \pi$ and $\alpha_0 = 1.5\ii$. The detector amplitude-efficiency $\eta$ and the channel transmissivity $T$ are varied. 
Fig. \ref{cat_states} shows the Schr\"odinger cat state before and after teleportation. The quantum character of the state is manifested in the negative value at the center of the Wigner function. Our results show that when the detector efficiency is taken into account ($\eta^2 =0.99$), the quantum character of the cat state is lost after 5dB of fixed attenuation ($T_A = T_B = 0.562$).

\subsection{Fading channel}
\label{fading_simulation}

\begin{figure}[h!]
		\centering
    \includegraphics[width=\linewidth]{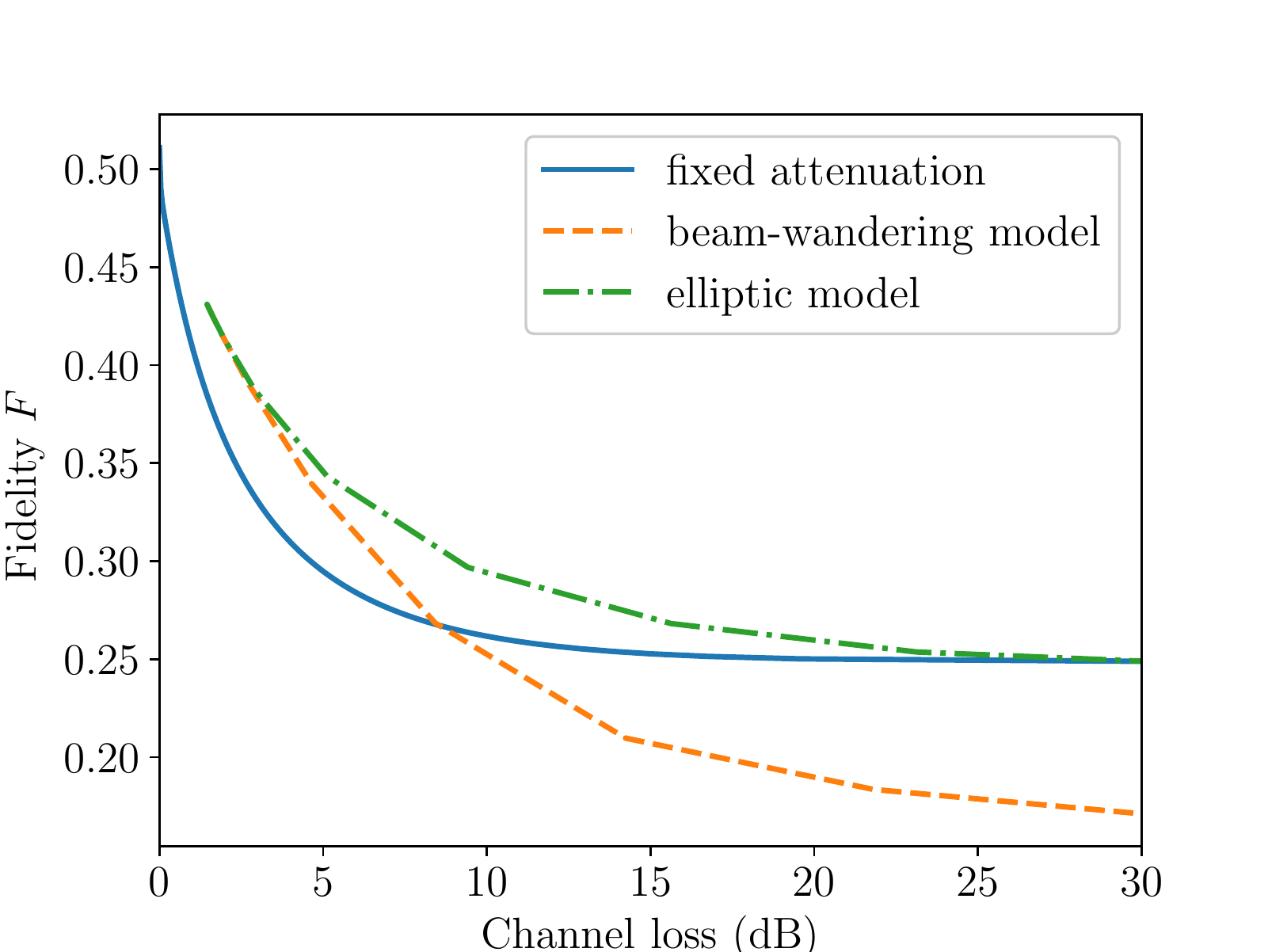}
		\caption{ The fidelity ($F$) of the Schr\"odinger's-cat state is plotted for different teleportation channels. 
		The blue line represents the teleportation fidelity of a fixed attenuation channel (plotted against the fixed channel loss). The other two lines represent a free-space fading channel with atmospheric turbulence (plotted against the mean channel loss). 
		The attenuation of the free-space channel is generated by different models: the beam-wandering model (in orange dashed line) and the elliptic model (in green dot-dashed line). (There is a cut-off loss that results from the cut-off transmissivity defined in Eqs. (\ref{t0_circular}) and (\ref{t_0_elliptic})).}
		\label{cat_fidelity}
\end{figure}

In this subsection, we assume a realistic scenario where the satellite is 500km high, the beam waist at the transmitter is $W_0 = 12$cm, and the wavelength is 780nm. In this setting, the beam width at the receiver is around $W=$1m. The receiver is designed to have an aperture $a=W=1$m. The variance of the beam center ($\sigma^2$) is the key parameter that influences the transmissivity of the channel. 

In Fig. \ref{circular_vs_elliptical_t_all_sigma_1e7}, the probability density $p(T)$ is plotted against $T$ for different values of $\sigma$, where higher $\sigma$ represents stronger atmospheric disturbance. The dashed lines show the probability densities generated by the beam-wandering model, while the solid lines show the probability densities generated by the elliptic model. 
In addition, our calculation shows that, in the beam-wandering model, the mean loss of 50\% (3dB) is reached when $\sigma = 0.7 a$, where $a$ is the aperture radius. On the other hand, for the elliptical model, 
the mean loss of 50\% (3dB) is reached earlier, when $\sigma = 0.4 a$.

\subsection{Fidelity}
The fidelity of the Schr\"odinger's-cat state after teleportation is calculated based on Eq. (\ref{fidelity}) and is plotted in Fig. \ref{cat_fidelity}. The different lines represent different attenuation models for the CV teleportation channel. The mean loss (in dB) is calculated by $-10 \log_{10}(T_A T_B)$ where $T_A$ and $T_B$ are the channel transmissivities. The fixed parameters are $r=1.15$, $\alpha_0 = 1.5\ii$, $\phi = \pi$ and $\eta^2 =0.99$. 

The solid blue line in Fig. \ref{cat_fidelity} represents a fixed attenuation channel. 
The dashed orange line and the dash-dotted green line represents free-space channels with atmospheric turbulence described by the beam-wandering and the elliptic model, respectively. $T_A$ and $T_B$ are generated by the simulation in subsection \ref{fading_simulation}, while the teleportation gain is set as $g = \sqrt{T_B/T_A}$. The output of the teleportation is calculated and compared to the input state. 
Our results show that when the mean loss of the fading channel is less than 30dB, which is the regime considered applicable for a satellite down-link channel from low-Earth-orbit, the fidelity of the cat state is higher than
that for a cat-state that has  passed through fixed attenuation channel of loss equal to the corresponding mean fading loss.

Furthermore, due to the different distributions of $T$ for the elliptical model and the beam-wandering model (see  Fig. \ref{circular_vs_elliptical_t_all_sigma_1e7}), these models possess
different probability distributions for the variance $V$ used in Eq. (\ref{general_teleportation}). This has the consequence that for a given mean loss the elliptical model leads to higher fidelities.
We have also carried out additional simulations for a wide range of parameter settings ($r$ from 0.5 to 3, $|\alpha_0|^2$ from 0.1 to 25, and $\phi$ from 0 to $\pi$) and find the main conclusions drawn here remain intact.

\section{Conclusion}
\label{conclusion}
In this work, we studied the effect of channel transmission loss on CV entanglement. Specifically, we used a lossy CV teleportation channel to teleport a Schr\"odinger's-cat state. Our results showed that when the channel has fixed attenuation (such as in an optical fiber), the quantum character of the cat state is lost after 5dB. We then investigated the scenario of an Earth-satellite channel with transmission loss induced by atmospheric turbulence. The attenuation in this realistic channel follows a probabilistic distribution which can be modeled by either the beam-wandering model or the elliptic model. Our results showed that for a down-link channel of less than 500km of length, i.e, less than 30dB of average loss, the teleported state attains higher fidelity than in a fixed attenuation channel. 
In reality, the beam profile after atmospheric turbulence is a hybrid of the beam-wandering and the elliptic models. While most past studies are based on the beam-wandering model, experimentalists should beware that the actual fidelity should be higher according to the elliptical model. 

\bibliographystyle{IEEEtran}
\bibliography{../myBib_short}

\end{document}